# A Raspberry Pi-based Traumatic Brain Injury Detection System for Single-Channel Electroencephalogram

**Navjodh Singh Dhillon [1], Agustinus Sutandi [1], Manoj Vishwanath [2], Miranda M. Lim [3,4], Hung Cao [2,5*], Dong Si [1*]**

[1] Computing and Software Systems, University of Washington, Bothell, Washington, USA; navjodh@uw.edu (N.D.); sutandia@uw.edu (A.S.); dongsi@uw.edu (D.S.)
[2] Department of Electrical Engineering and Computer Science, University of California, Irvine, California, USA; manojv@uci.edu (M.V.); hungcao@uci.edu (H.C.)
[3] VA Portland Health Care System, Portland, Oregon, USA; lmir@ohsu.edu
[4] Department of Neurology, Oregon Health & Science University, Portland, Oregon, USA
[5] Department of Biomedical Engineering, University of California, Irvine, California, USA
* Correspondence: hungcao@uci.edu (H.C.); dongsi@uw.edu (D.S.)

**Abstract:** Traumatic Brain Injury (TBI) is a common cause of death and disability. However, existing tools for TBI diagnosis are either subjective or require extensive clinical setup and expertise. The increasing affordability and reduction in the size of relatively high-performance computing systems combined with promising results from TBI related machine learning research make it possible to create compact and portable systems for early detection of TBI. This work describes a Raspberry Pi based portable, real-time data acquisition, and automated processing system that uses machine learning to efficiently identify TBI and automatically score sleep stages from a single-channel Electroencephalogram (EEG) signal. We discuss the design, implementation, and verification of the system that can digitize the EEG signal using an Analog to Digital Converter (ADC) and perform real-time signal classification to detect the presence of mild TBI (mTBI). We utilize Convolutional Neural Networks (CNN) and XGBoost based predictive models to evaluate the performance and demonstrate the versatility of the system to operate with multiple types of predictive models. We achieve a peak classification accuracy of more than 90% with a classification time of less than 1 s across 16 s - 64 s epochs for TBI vs control conditions. This work can enable the development of systems suitable for field use without requiring specialized medical equipment for early TBI detection applications and TBI research. Further, this work opens avenues to implement connected, real-time TBI related health and wellness monitoring systems.

## 1. Introduction

### *1.1. Background*

Traumatic Brain Injury (TBI) is a form of acquired brain injury caused by external impact to the head that results in damage to the brain [1]. It is a common cause of death and disability in the United States (U.S.) and can be caused by a variety of factors including falls, motor vehicle crashes, sports, or combat injuries. TBI affects an estimated 2 million people in the U.S. across all age groups, according to data from U.S. Centers for Disease Control and Prevention (CDC) [2] and likely a larger number globally. TBI often leads to neurological problems in individuals, including cognitive, motor, and sleep-wake dysfunction [3].

Currently available medical tools for TBI diagnosis are largely subjective [4] and a lack of consensus regarding what constitutes mild TBI (mTBI) adds to the complication of the under-diagnosis of mTBI [5]. TBI is categorized into mild, moderate, or severe based on the Glasgow coma scale (GCS), Loss of consciousness (LOC), and Post-traumatic amnesia (PTA) [6] which are qualitative tests rather than quantitative measures. The World Health Organization's (WHO) definition of mTBI allows for a GCS score of 13–15 to be assessed after the typical 30-min timeframe, which accounts for the expected time of arrival of a qualified healthcare provider [7]. However, the GCS has its drawbacks. Being highly inter-observer dependent makes it necessary to report exact findings rather than just the score. In addition, one of the key parameters in GCS is the eye score which might be unattainable in case of an eye injury.

Considering the requirements from a medical resource standpoint, existing clinical tools used to diagnose mTBI such as Magnetic Resonance Imaging (MRI) and Computer Tomography (CT) [4] require an extensive, high-cost clinical setup and specialized operator skill set which are not always available at the time and place of an incident, and these neuroimaging tests may still be negative in many cases of mTBI. As a result of the limitations of the present-day methods used to detect TBI, there is a need for new technology capable of rapid, accurate, non-invasive, and most importantly, field-capable detection of mTBI to bridge the technological gap that exists today. Early, objective, and reliable mTBI detection can help affected individuals undergo timely monitoring and therapy and can prevent death in severe cases.

Machine learning techniques provide a way to study mTBI and create systems to help objectively diagnose and monitor mTBI presence and stages in individuals. Recently, machine learning techniques have been investigated for the purpose of classifying mTBI from electroencephalogram (EEG) data in mice based on models created using the lateral Fluid Percussion Injury (FPI) method [3]. FPI induced mice demonstrate very similar behavioral deficits and pathology to those found in humans afflicted with mTBI, including sleep disturbances [8, 3]. In this work, we use EEG data acquired from the compelling FPI mouse model of mTBI. Previous investigations have studied a variety of classification techniques, including classical machine learning such as SVM [9], and deep learning such as Convolutional Neural Networks (CNN) [9, 10]. These techniques have been shown to perform TBI classification with more than 80% accuracy [9]. However, in most investigations we reviewed (as described in the subsequent Related Works subsection) that implement machine learning for TBI detection, the primary focus was the study of classification techniques and performance of classification models [9] rather than portable deployment. In cases, where portable deployment was involved, the focus was application-specific implementation, for example, closed-loop robotic control systems [11].

This work focuses on creating a machine learning-based, fast, portable, and ready to use EEG classification system for mTBI detection using Raspberry Pi 4 (RPi). This system works with a variety of machine learning models and can be used along with live EEG recording systems to detect mTBI. This capability can enable field use and make early mTBI detection possible without the requirement of extensive medical setup or specialized medical domain knowledge. Further, this work has the potential to create avenues for implementing mTBI related real-time connected health monitoring systems [12] and allow further research on real-time mTBI detection using EEG.

The deployment system created in this work incorporates an Analog to Digital Converter (ADC) front-end and utilizes Convolutional Neural Network (CNN) and XGBoost [13] predictive models to perform sleep staging and detect the presence of mTBI using a single-channel EEG signal. The system captures and classifies EEG epochs into four target classes – Sham Wake, Sham Sleep, mTBI Wake, and mTBI Sleep. We demonstrate that our system can capture physical EEG signals and perform feature extraction and prediction using the XGBoost model in the order of 0.02 s per epoch which makes it possible to quickly detect the presence of mTBI. We also verify that the cross-validation metrics obtained on the RPi based system are identical to those obtained on a High-Performance Computer (HPC) such as a 64-bit workstation computer running macOS or Windows.

The Related Works subsection covers previous relevant investigations in this area. We describe the deployment system design, operation, classification model configuration, and validation techniques in the Methods section. The Results section covers the performance comparison of XGBoost with CNN and the deployment system performance evaluation. Finally, we conclude our work and discuss possible future directions in the Conclusions section.

### 1.2. Related Works

The study of brain activity using electroencephalogram (EEG) typically involves extracting information from signals associated with certain activities. In recent years, machine learning techniques have been applied to the classification of mTBI because it enables the extraction of complex and typically non-linear patterns from the EEG data [14–17]. Most of the work surveyed used rule-based techniques, such as k-Nearest Neighbors (k-NN).

A few systems used a small, portable computer for deployment in some form. The Neuroberry platform [18] used a Raspberry Pi 2 device to capture EEG signals but the focus was on enabling EEG signal availability on the Internet of Things (IoT) domain. The Acute Ischemic Stroke Identification System [19] utilized an Analog to Digital Converter (ADC) front end with Raspberry Pi 3 to capture physical EEG signals. However, this system transferred the captured data to an HPC running MATLAB for signal analysis and processing and did not focus on signal classification. Zgallai *et al* [11] described a Raspberry Pi-based system that used deep learning to perform EEG signal classification. It was designed to identify a subject's intended movement direction from a multi-channel EEG signal to control wheel-chair movement in a closed-loop robotic system rather than as a general system for identification, analysis, and monitoring of a physiological condition such as mTBI. Bruno *et al* [20] highlighted challenges with existing medical diagnosis techniques and described a classification system from the perspective of real-time TBI diagnosis, but their work was focused on the algorithm to perform TBI diagnosis and not on the implementation of a deployment system.

In our previous work [10], we developed and described a CNN based model to perform automated sleep stage scoring and mTBI classification. In addition, we did a limited deployment of the CNN model on a Raspberry Pi 4 system. In that work, the focus was on describing the CNN model configuration, evaluating its performance, and showcasing that deployment to RPi was feasible rather than designing a complete, portable classification system. We have reused the previously developed CNN model in the current work to provide a baseline performance comparison with a new XGBoost model developed for this work. Further, the two models enable us to demonstrate the versatility of the current system to operate with multiple types of predictive models.

To the best of our knowledge, no standalone, portable system has yet been created using Raspberry Pi that can capture real-time EEG signals, detect the presence of mTBI, and classify mTBI sleep/wake epoch states.

## 2. Materials and Methods

In this section, we describe in detail the dataset used, software and hardware design, and operation of the RPi based classification system. We also describe the classification model configuration and criteria used for system verification.

### 2.1. Dataset Details

A previously published dataset as described in [3] was used to train and evaluate deployed models. This dataset was collected as part of a study involving 11 adult male mice subjects divided into two groups – mTBI and Sham. FPI [8] procedure was used to induce mTBI in 5 subjects and the remaining 6 mice were used as Sham (control) subjects. Single-channel EEG data were captured from each subject at a 256 Hz sampling rate over 24 hours. Sleep stages were scored manually by experienced scorers into 4 s epochs of wake, non-rapid-eye movement (NREM), and rapid-eye-movement (REM) stages. In this work, we combined NREM and REM sleep stages in the dataset to a single sleep stage, resulting in 4 target classification classes - Sham Wake, Sham Sleep, mTBI Wake, and mTBI Sleep [10].

## 2.2. System Design

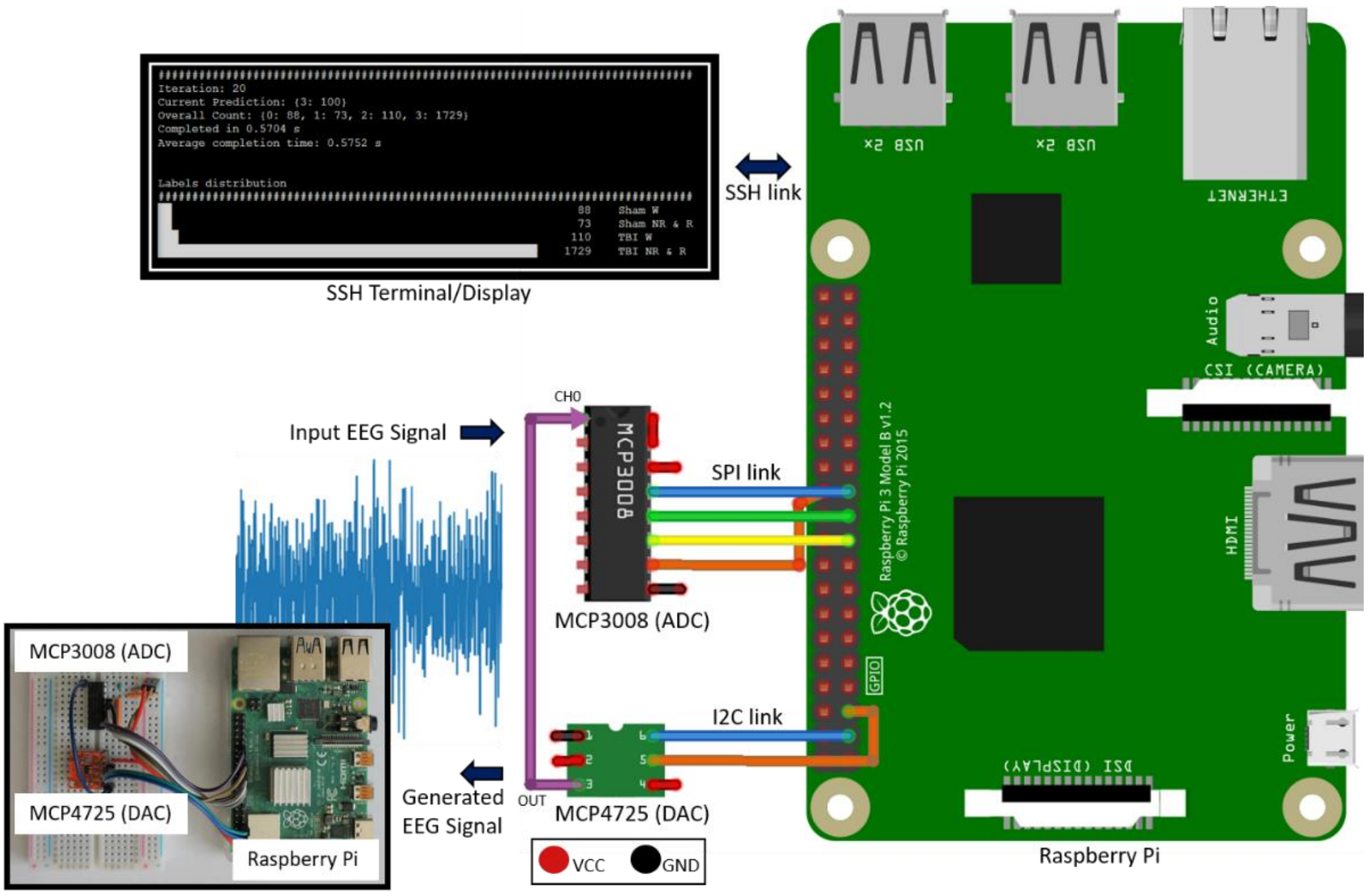

**Figure 1.** Deployment system hardware setup.

The hardware design of the system is illustrated in Figure 1. MCP3008 [21] Analog to Digital Converter (ADC) is used as the hardware front-end to capture and digitize input EEG signals. It is a 10-bit ADC with Serial Peripheral Interface (SPI) that connects to RPi's SPI interface. While MCP3008 is suitable for single-channel EEG applications like our work, having 8 channels per Integrated Circuit (IC) makes it possible to expand the current system to a multi-channel configuration, if needed. It is a low power device with 5 nA standby current and 500 μA operating current, making it suitable for standalone, portable embedded system applications. It supports a data rate of 200 kilo-samples per second (ksps), which provides sufficient data capture and transfer speed for typical EEG sampling rates, 256 Hz in this case. Also, a 200 ksps peak sampling rate exceeds the basic Nyquist criteria (1) and allows us to sample at more than 10 times the peak signal frequency component (0.5 Hz - 60 Hz for the current signal) to fully capture the frequency, amplitude, and shape of EEG signals.

$$f_s \geq 2f_c \tag{1}$$

where $f_s$ is the sampling frequency and $f_c$ is the highest frequency component in the signal.

MCP4725 [22] is used to generate the EEG time series signal voltage levels from an EEG data file. MCP4725 is a 12-bit Digital to Analog Converter (DAC) with Inter-Integrated Circuit (I2C) interface that is suitable for connectivity with RPi's I2C host interface. Having the ability to generate a physical EEG signal from a raw EEG data file makes it possible to validate the system operation without requiring a live signal from a subject.

To view live classification results and for initiating and controlling system operation, either a dedicated touch display or an ethernet Secure Shell (SSH) connection can be used. For this work, we used an SSH connection.

## 2.3. Classification System

Figure 2 shows the architecture of the classification system. The input was a single-channel EEG signal sampled at 256 Hz. Two different machine learning model implementations were used for deployment on RPi – CNN and XGBoost.

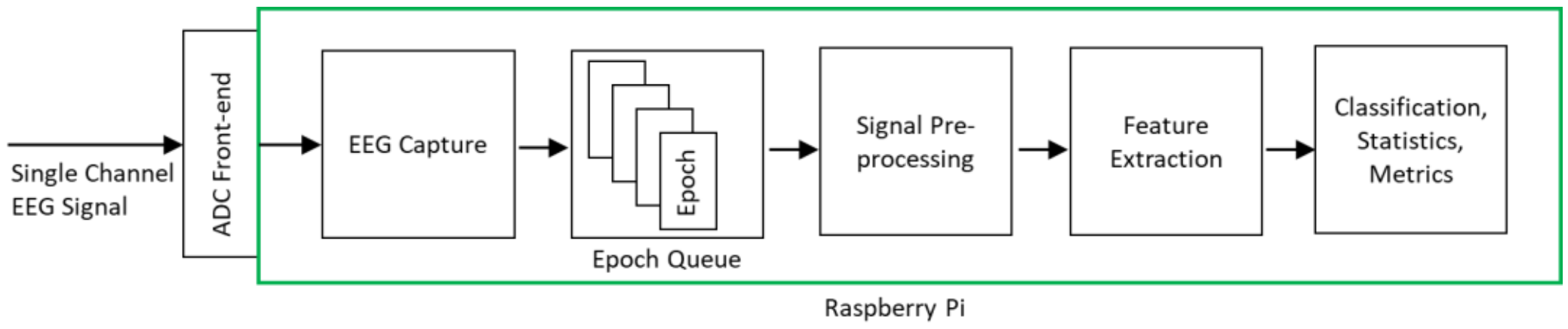

**Figure 2.** Proposed system architecture.

The first implementation used a CNN model [10] to automatically extract features suitable for classification from an EEG signal with an epoch duration of 16 s to 64 s. The second implementation involved XGBoost that also used a 16 s to 64 s epoch duration. In this case, decibel normalized sub-band powers and ratio of theta to alpha sub-band power were extracted from the EEG signal as described in [23]. The extracted features were fed to an XGBoost classifier to obtain the predicted classes. An epoch size of 16 s to 64 s is optimal as it allows sufficient distinguishing patterns to be captured from the signal for accurate and reliable classification, and at the same time is small enough for fast prediction and deployment to an embedded device with limited memory and processing resources such as RPi.

*2.4. System Operation*

EEG time series data points were loaded into RPi memory from the EEG data file and were then transferred to the MCP4725 DAC for generation, thus recreating the time series EEG signal. The generated EEG signal was captured by the MCP3008 ADC sampling at 256 samples per second. The captured EEG data points were transferred to the RPi system and accumulated into epoch buckets. The EEG epochs were then placed on the processing queue of the software processing system to undergo preprocessing, feature extraction, and classification stages.

To enable the system to operate in real-time capacity without losing EEG signal data points, we used a queue-based software design with separate threads for EEG signal capture and processing. This ensured incoming EEG data points could be collected while previously collected EEG data points were being processed. Figure 2 illustrates the queue-based processing system operation.

The implemented system was designed to display the inferred label count, epoch processing time, and distribution of epoch count per classified label via a histogram (Figure 3). The capability of the system to provide a live view of inferred EEG epoch labels distribution enabled early inference of whether a subject is afflicted with mTBI or not.

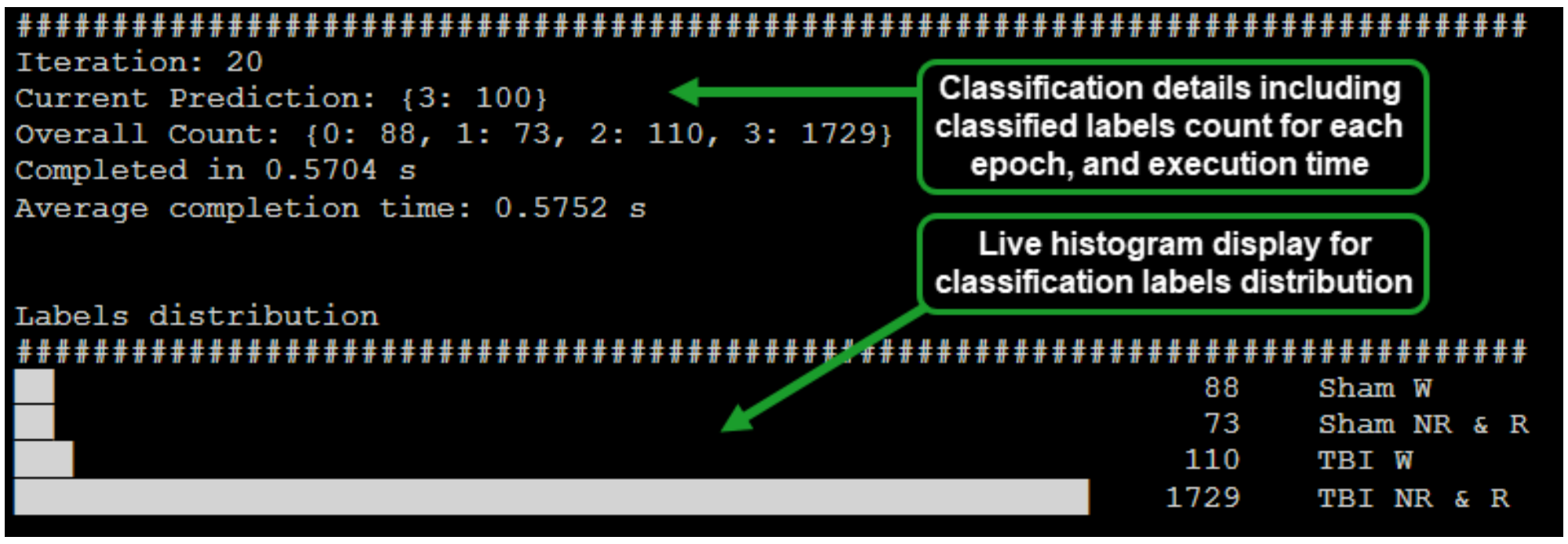

**Figure 3.** Live classification display for the RPi based system (64 s epoch, batch size: 100).

*2.5. System Verification*

This section describes the validation techniques and metrics applied to the deployment system to consider it successful for its intended purpose. The actual verification results are covered in the Results section.

The EEG data used in this work comprised of 11 mice divided into two groups – mTBI and Sham. Labeled non-overlapping epochs were created from the 24-hour recording to train and evaluate the classification models across 10 folds and mean accuracy values across all folds were calculated. The current dataset offers limited generality because of the small number of subjects and hence we used random sampling to split non-overlapping epochs across all subjects into training and testing datasets. The current work focuses on describing the live classification system design and operation, and from a generality perspective, the accuracy results in the current work should be validated with a separate, larger dataset [10].

To evaluate the RPi based deployment system functionality, classification metrics and inference time were calculated and compared to that of an HPC. The operation of the queue-based system architecture was evaluated by calculating the inference time and processing time per epoch batch with different batch sizes. We also verified that the generated EEG epochs were captured and processed by the queue-based system without data loss by tracking the generated and captured epoch count. The classification performance of the deployment system was evaluated using accuracy, precision, and recall as metrics. The definitions of these metrics are as follows:

$$accuracy = \frac{TP + TN}{TP + TN + FP + FN} \quad (2)$$

$$precision = \frac{TP}{TP + FP} \quad (3)$$

$$recall = \frac{TP}{TP + FN} \quad (4)$$

where TP: True Positives, TN: True Negatives, FP: False Positives, and FN: False Negatives.

CNN and XGBoost models were trained on an HPC and deployed on the RPi system to calculate metrics and to perform live classification. We evaluated the timing performance of the live operation of the system and verified that the classification model provided valid output labels for the input EEG epochs. The actual timing and classification metrics are provided in the Results section.

We verified that the signal generated on the RPi system using the DAC and captured by the ADC was consistent with that stored in the EEG signal data file. To compare the stored and generated signals, and as a measure of generated signal quality, we used the Mean Squared Error (MSE) value as a metric, which is calculated as follows:

$$\left(\frac{1}{n}\right) * \sum_{i=1}^{n} \left(Y_i - \hat{Y}_i\right)^2 \quad (5)$$

where $n$ is the number of data points, $Y_i$ is the observed set of magnitude values generated by the DAC and $\hat{Y}_i$ is the expected set of magnitude values in the EEG data file.

## 3. Results

### 3.1. Performance Comparison of RPi with HPC

The CNN and XGBoost predictive models trained on the HPC were deployed on RPi to verify system functionality, retrieve classification metrics, and calculate inference time. Accuracy, precision, and recall results on RPi were identical to those achieved with an HPC across epoch lengths varying from 16 s to 64 s, as shown in Table 1.

**Table 1.** System performance comparison of RPi with HPC with 4 classes using XGBoost.

| Device | PC | RPi | PC | RPi | PC | RPi |
|---|---|---|---|---|---|---|
| Epoch | 16 | 16 | 32 | 32 | 64 | 64 |
| Accuracy | 0.982 | 0.982 | 0.974 | 0.974 | 0.968 | 0.968 |
| Sham Wake | | | | | | |
| Precision | 0.972 | 0.972 | 0.981 | 0.981 | 0.981 | 0.981 |
| Recall | 0.986 | 0.986 | 0.986 | 0.986 | 0.982 | 0.982 |
| Sham Sleep | | | | | | |
| Precision | 0.973 | 0.973 | 0.951 | 0.951 | 0.937 | 0.937 |
| Recall | 0.951 | 0.951 | 0.945 | 0.945 | 0.948 | 0.948 |
| mTBI Wake | | | | | | |
| Precision | 0.989 | 0.989 | 0.961 | 0.961 | 0.951 | 0.951 |
| Recall | 0.990 | 0.990 | 0.961 | 0.961 | 0.934 | 0.934 |
| mTBI Sleep | | | | | | |
| Precision | 0.998 | 0.998 | 0.997 | 0.997 | 0.989 | 0.989 |
| Recall | 0.997 | 0.997 | 0.995 | 0.995 | 0.990 | 0.990 |

### *3.2. Epoch Processing Time*

The variation of epoch processing time (which includes preprocessing, feature extraction, and classification operations) with the number of epochs is shown in Figure 4. We observed that the processing time on RPi increased with an increase in the number of epochs processed in a batch. This was expected as the epochs are processed sequentially and the processing time per epoch is accumulated. We discuss its significance in the subsequent Discussion section.

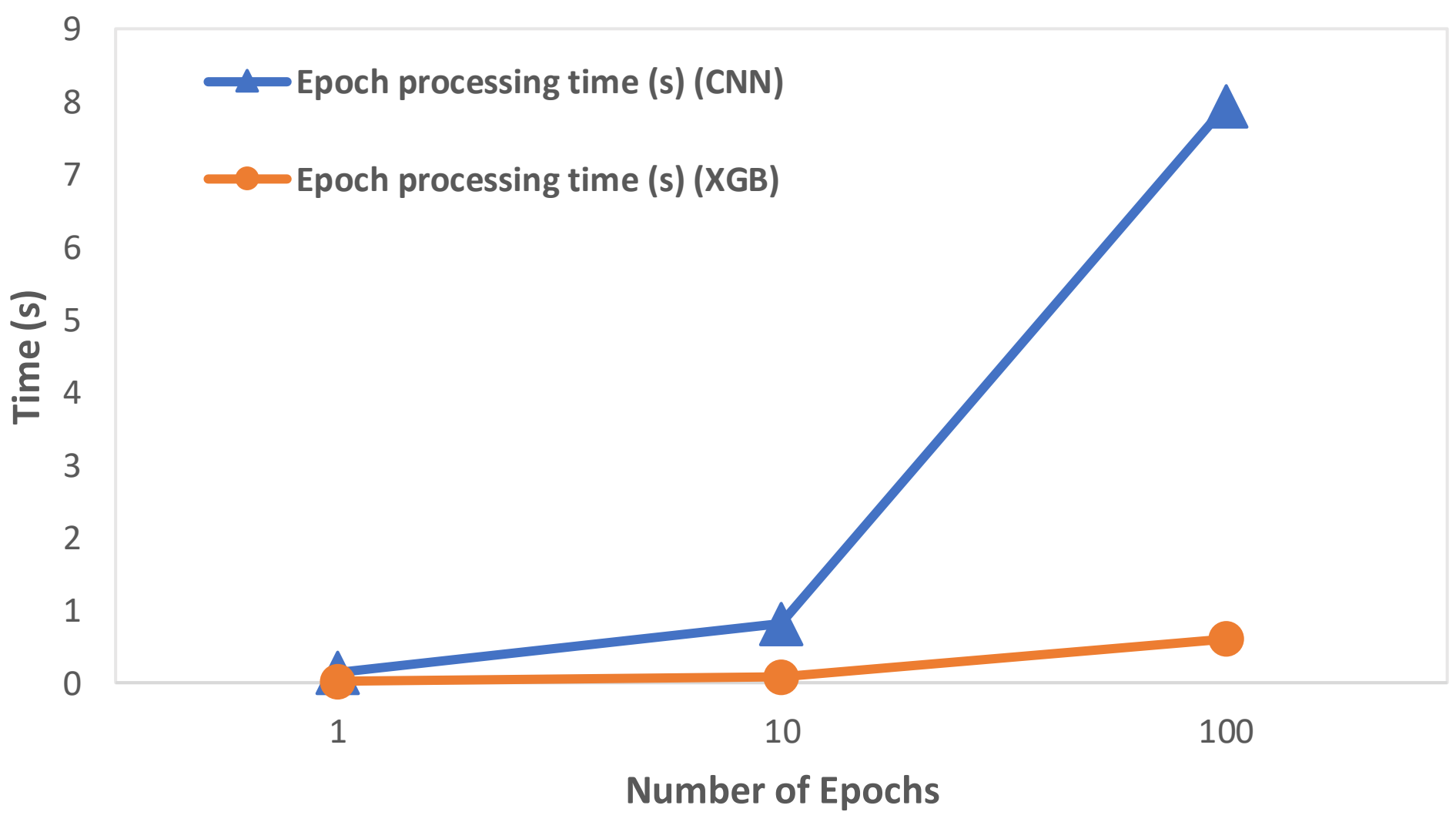

**Figure 4.** Variation of epoch processing time with the number of epochs on RPi (64 s epoch).

### *3.3. Generated EEG signal quality*

MSE was used as a metric to compare stored and generated EEG waveforms, as a measure of generated signal quality. We observed a typical MSE value of 0.26, which indicated high fidelity between the DAC generated signal and stored signal data points.

### *3.4. Performance Comparison of CNN and XGBoost*

We compared classification metrics and performance of the XGBoost and CNN models on the deployment system as well as an HPC. Figure 5 shows the variation of accuracy with change in epoch size and Figure 6 shows the variation of inference time per epoch for the XGBoost (labeled as XGB) and CNN models. Accuracy across XGBoost and CNN was found to decrease slightly (0.01%) with each increase in epoch size. Compared to CNN, the overall accuracy for XGBoost was better by 12 to 15 percentage points across various epoch sizes. The inference time for XGBoost was found to be significantly faster than CNN, especially for execution on RPi. For a 64 s epoch size, the inference time of XGBoost was about 0.004% of the classification time of CNN on RPi. For XGBoost, we found that the inference time per epoch for RPi was within 1 µs of that on an HPC. Further, the variation of inference time remained roughly within 2 µs with each increase or decrease of epoch length for both HPC and RPi. Overall, the timing performance and accuracy were found to be better in the case of XGBoost compared to CNN, which would render XGBoost more suitable for use in a deployment configuration on RPi.

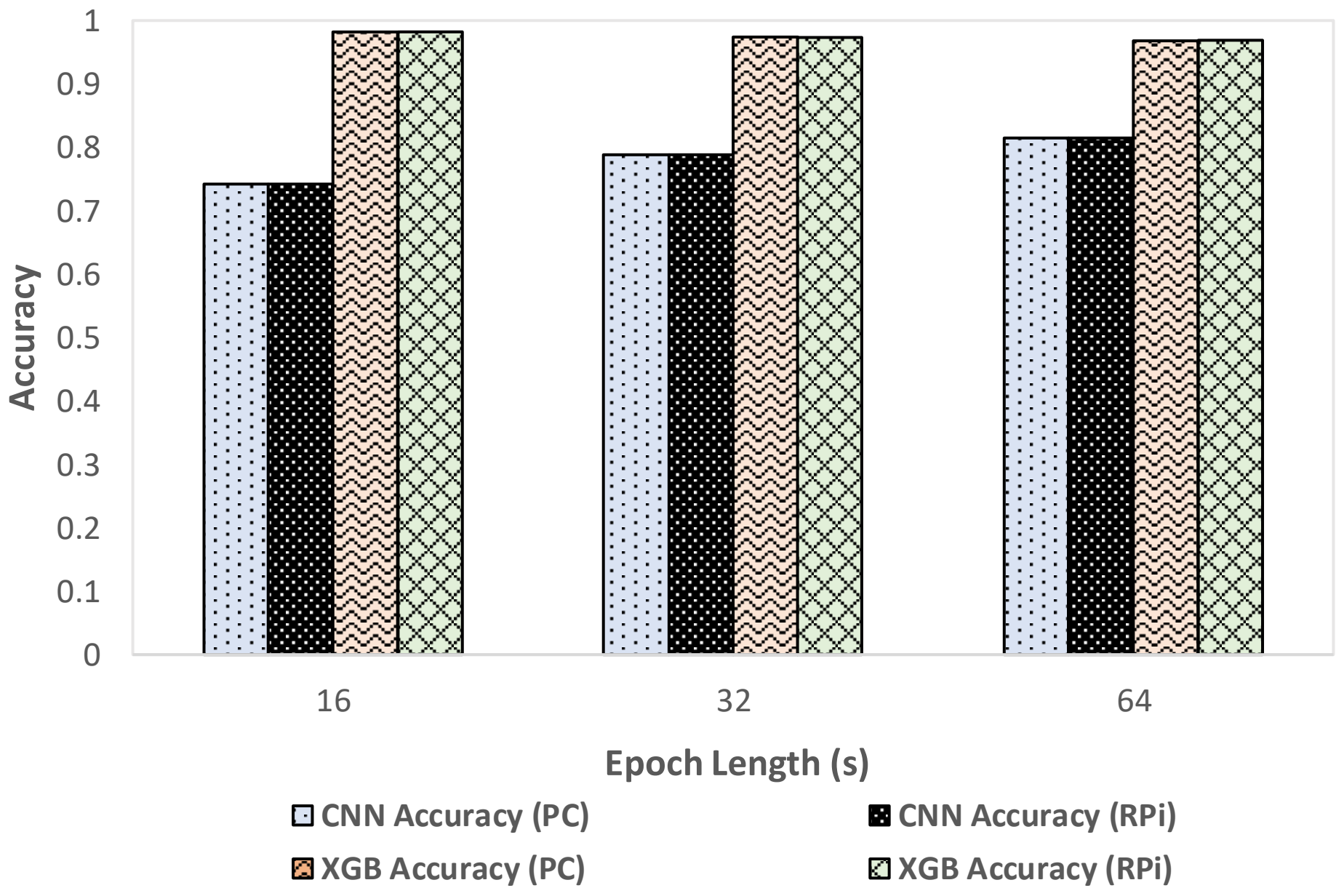

**Figure 5.** Variation of accuracy with epoch size.

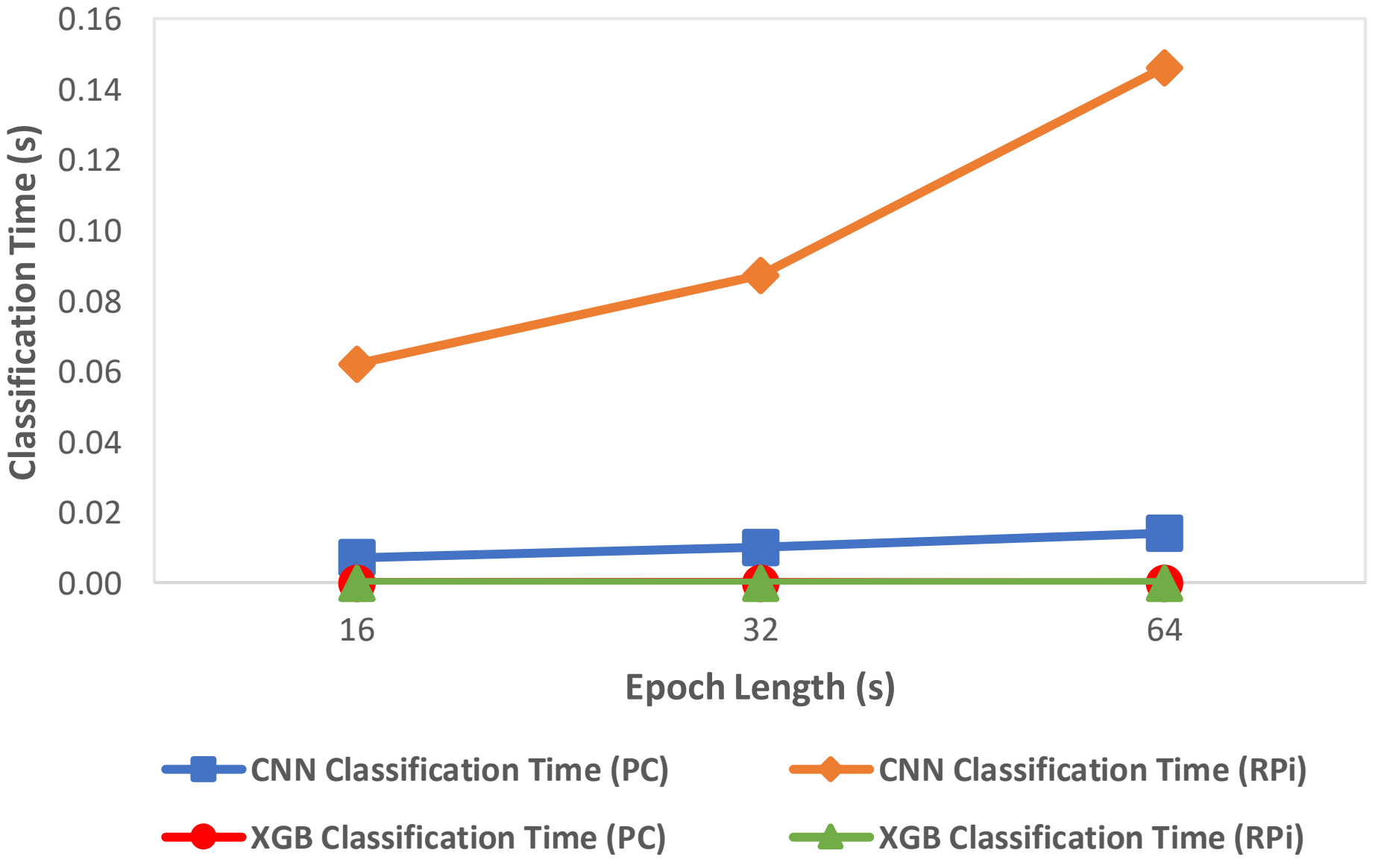

**Figure 6.** Variation of classification time with epoch length.

### 3.5. Queue-based Processing System Validation

We verified the functionality of the queue-based processing system on RPi by generating 100 epochs of 64 s duration and capturing those epochs in the processing loop. We found that all generated epochs were consistently captured and processed, resulting in a 0% epoch loss across 5 test runs. Details on timing related to EEG epoch collection and processing are shown in Table 2. The processing time (which includes preprocessing, feature extraction, and classification operations) ranged from 0.01% to 0.03% of epoch collection time. This meant the system could process a given number of epochs significantly faster than the time it took for the system to collect the epochs.

**Table 2.** Epoch collection and processing time (64 s epoch).

| Number of Epochs | Epoch collection time (s) | Epoch processing time (s) | Processing time as % of collection time |
|---|---|---|---|
| 1 | 64 | 0.02 | 0.03% |
| 10 | 640 | 0.08 | 0.01% |
| 100 | 6400 | 0.6 | 0.01% |

## 4. Discussion

In this work, we proposed and demonstrated an RPi based EEG acquisition, processing, and classification system for early mTBI detection and sleep stage classification. This system was demonstrated to operate in a portable, real-time, and standalone configuration and perform classification of real-time EEG epochs into four target classes (sham wake, sham sleep, mTBI wake, mTBI sleep).

As shown in Table 1, the accuracy, precision, and recall results were identical across RPi and HPC. This confirmed that the predictive model behavior did not change when the training and deployment systems involved different system architectures, *i.e.*, x64 based MacOS/Windows HPC for training vs. ARM-based RPi for deployment and prediction. Hence, it is possible to train a predictive model on a more powerful computer (HPC) and deploy it to an embedded device such as RPi that has limited memory and processing resources. This is especially applicable to multi-layered neural networks like CNN that typically have long training times on an HPC, and the training times would be prohibitively long on an embedded device like RPi.

We calculated the epoch processing time (which included preprocessing, feature extraction, and classification operations) on RPi by varying the number of epochs, as shown in Figure 4 and described in Table 2. While it was expected that the processing time would increase as the number of processed epochs is increased, the key inference was that the processing time was significantly smaller than the time required to collect the EEG epochs. At 256 Hz sampling rate and 64 s epoch size, the processing time ranged from 0.01% to 0.03% of the epoch collection time. Hence, we concluded that the system had ample time to process previously captured EEG epochs while new epochs were captured at practical EEG signal sampling rates.

We employed two different approaches for supervised learning models used in this system, the CNN model developed in our previous work, and an XGBoost predictive model created in the current work. We compared classification metrics and performance of the XGBoost and CNN models on the deployment system as well as an HPC. We observed that the XGBoost model exhibited significant performance improvement in terms of accuracy (as shown in Figure 5,) and inference time (as shown in Figure 6) compared to the CNN based predictive model. In the case of XGBoost, the variation of inference time remained roughly within 2 μs between HPC and RPi. A low inference time was critical for the real-time operation of the classification system. One possible reason for the better accuracy performance in the case of XGBoost compared to CNN was that the classification model for XGBoost was created using hand-crafted features which enabled learning differentiating patterns for the four target classes better than the CNN model that automatically extracted the differentiating features. These results, however, were data-dependent, so they should be validated on different datasets to verify the generality of the model. We found that overall, XGBoost was better suited for deployment on RPi because of its faster inference time and better performance than CNN. By using two different predictive models for classification, we demonstrated the flexibility of the system to deploy improved classification models in the future.

In this system, we used a DAC to generate EEG signal waveform form European data format (EDF) files. This provided a reliable way to generate an EEG signal waveform without requiring an actual subject to capture the EEG signal from. We verified that the EEG waveform generated using the DAC on RPi was consistent with the EEG data stored in the EDF file. The verification was done by calculating MSE across the stored and generated signal, which was found to be 0.26, a small value indicating that the generated signal represented the stored signal accurately. Synthesizing EEG signals to replicate the complex and typically non-linear signal patterns is challenging and the ability to generate EEG signals from an actual recording data file using a DAC simplifies the setup that is required to test an EEG classification deployment system hardware and software chain. It enables the use of several available open-access EEG data files to train classification models and test the deployment system. For future use, the signal generation capability of this system can be simplified for ease of use and expanded to work with a variety of EEG data file types. This can help accelerate mTBI related future research pertaining to portable classification systems that are often constrained by the lack of readily available live EEG signals to test a hardware classification system.

In addition to early mTBI detection, the capability of the system to perform live classification on input EEG signals can be extended to cover mTBI related health and sleep monitoring applications in the future. Typically, after the initial diagnosis, TBI patients undergo EEG sleep monitoring in a hospital setup. A portable EEG sleep monitoring system, such as the one described in this work, can enable a subject to self-monitor in home settings, greatly enhances the accuracy, efficiency, and efficacy.

The classification system developed in the current work can also provide a replacement of the labor-intensive manual sleep-stage scoring of EEG signals by human experts with an online and automated system with the capability to perform fast sleep staging. Further, our technical approaches can be extended to several other EEG applications, including detection of the onset of epileptic seizures, strokes, and other neurological conditions.

In this work, we used a relatively simple hardware system to capture and digitize EEG signals, which could be improved. Because we generated EEG signals from a data file containing clean EEG data, this hardware did not include amplification and filtering stages. A practical system designed for field use would require additional hardware and software capabilities to capture and process EEG signals in real-time. In terms of hardware, such a system would require amplification, preprocessing, and filtering stages. In software, decimation, normalization, Independent Components Analysis (ICA), physiological artifact removal (e.g., eye and muscle movement artifacts), and filtering stages can be implemented. Further, we used an 8-bit ADC for this proof-of-concept system, but for devices designed for practical use, ADCs typically vary from 16-bit to 24-bit resolution. For example, the OpenBCI Cyton Biosensing system [24] for sampling EEG and other physiological signals uses a 24-bit ADC. We will note that higher resolution ADCs also involve a relatively higher cost and have lower sampling rates as the number of resolution bits increases. Also, the system in this work was designed for single-channel EEG generation and capture, which limits its use for multichannel EEG applications. The current system also assumes a direct single-channel EEG electrode connection to the ADC input. It does not directly provide connectivity to wireless (Bluetooth and Radio Frequency) EEG headsets. However, several 'hardware attached on top' (HAT) devices are available for RPi, for example, the brainHAT [25], that makes it possible to connect wireless headsets seamlessly and we anticipate the system in this work to function as intended with the actual streaming EEG data outside the particulars of EEG headset interfacing.

## 5. Conclusion

To the best of our knowledge, this is the first system capable of performing mTBI related EEG signal classification as described previously, deployed on a portable, low-cost device like RPi and designed to operate in a live configuration. The techniques developed in this work are general and can potentially be extended to create and deploy predictive models from EEG as well as other physiological signals acquired from human subjects, enabling e-care, self-care, and telemedicine.

**Acknowledgments:** This research was funded by the Graduate Research Award of Computing and Software Systems division and the startup fund 74-0525 of the University of Washington Bothell.

The data in this work was supported with resources and the use of facilities at the VA Portland Health Care System, VA Career Development Award #1K2 BX002712 to M.M.L. Interpretations and conclusions are those of the authors and do not represent the views of the U.S. Department of Veterans Affairs or the United States Government.